\documentclass[twocolumn,twocolumn]{IEEEtran}
\usepackage[T1]{fontenc}
\usepackage[latin9]{inputenc}
\usepackage{color}
\usepackage{float}
\usepackage{graphicx}
\usepackage[unicode=true,
 bookmarks=true,bookmarksnumbered=true,bookmarksopen=true,bookmarksopenlevel=1,
 breaklinks=false,pdfborder={0 0 0},pdfborderstyle={},backref=false,colorlinks=false]
 {hyperref}
\hypersetup{pdftitle={Your Title},
 pdfauthor={Your Name},
 pdfpagelayout=OneColumn, pdfnewwindow=true, pdfstartview=XYZ, plainpages=false}
\usepackage{breakurl}

\makeatletter

\floatstyle{ruled}
\newfloat{algorithm}{tbp}{loa}
\providecommand{\algorithmname}{Algorithm}
\floatname{algorithm}{\protect\algorithmname}

\usepackage[caption=false,font=footnotesize]{subfig}
\usepackage{algorithm}
\usepackage{algorithmic}

\usepackage{multirow} 
\usepackage{amsmath} 
\usepackage{xcolor}

\allowdisplaybreaks[4]

\ifCLASSOPTIONcompsoc
\usepackage[caption=false,font=normalsize,labelfont=sf,textfont=sf]{subfig}
\else
\usepackage[caption=false,font=footnotesize]{subfig}
\fi

\usepackage{cite}
\usepackage{bm}
\usepackage{algorithmic}
\usepackage{algorithm}
\usepackage{graphicx}
\IEEEoverridecommandlockouts

\usepackage{lettrine}

\usepackage{geometry}
\@ifundefined{showcaptionsetup}{}{%
 \PassOptionsToPackage{caption=false}{subfig}}
\usepackage{subfig}
\makeatother

\begin{document}

\title{\textcolor{black}{Recurrent LSTM-based UAV Trajectory Prediction
with ADS-B Information}}

\author{\IEEEauthorblockN{Yifan Zhang$^{\ast}$, Ziye Jia$^{\ast}$, Chao
Dong$^{\ast\mathparagraph}$, Yuntian Liu$^{\ast}$, Lei Zhang$^{\ast}$,
and Qihui Wu$^{\ast}$\\
 }\IEEEauthorblockA{$^{\ast}$The Key Laboratory of Dynamic Cognitive
System of Electromagnetic Spectrum Space,\\
Ministry of Industry and Information Technology, Nanjing University
of Aeronautics and Astronautics\\
$^{\mathparagraph}$Corresponding author, email: dch@nuaa.edu.cn}\thanks{}}

\maketitle
\pagestyle{empty} 

\thispagestyle{empty}
\begin{abstract}
Recently, unmanned aerial vehicles (UAVs) are gathering increasing
attentions from both the academia and industry. The ever-growing number
of UAV brings challenges for air traffic control (ATC), and thus trajectory
prediction plays a vital role in ATC, especially for avoiding collisions
among UAVs. However, the dynamic flight of UAV aggravates the complexity
of trajectory prediction. Different with civil aviation aircrafts,
the most intractable difficulty for UAV trajectory prediction depends
on acquiring effective location information. Fortunately, the automatic
dependent surveillance-broadcast (ADS-B) is an effective technique
to help obtain positioning information. It is widely used in the civil
aviation aircraft, due to its high data update frequency and low cost
of corresponding ground stations construction. Hence, in this work,
we consider leveraging ADS-B to help UAV trajectory prediction. However,
with the ADS-B information for a UAV, it still lacks efficient mechanism
to predict the UAV trajectory. It is noted that the recurrent neural
network (RNN) is available for the UAV trajectory prediction, in which
the long short-term memory (LSTM) is specialized in dealing with the
time-series data. As above, in this work, we design a system of UAV
trajectory prediction with the ADS-B information, and propose the
recurrent LSTM (RLSTM) based algorithm to achieve the accurate prediction.
Finally, extensive simulations are conducted by Python to evaluate
the proposed algorithms, and the results show that the average trajectory
prediction error is satisfied, which is in line with expectations.
\end{abstract}


\newcommand{\CLASSINPUTtoptextmargin}{0.8in}

\newcommand{\CLASSINPUTbottomtextmargin}{1in}

\section{Introduction}

\lettrine[lines=2]{T}{he} unmanned aerial vehicles (UAVs) are gathering
attentions from both the academia and industry, and are widely accepted
in various applications \cite{key-1} \cite{key-2} \cite{key-3}.
However, with the increment of UAVs, the low-altitude airspace becomes
extremely crowded. Therefore, it is imperative to deal with collisions
among  UAVs \cite{key-4}. Besides, UAVs work in accordance with their
pre-set flight routes, which means that they are unable to achieve
real-time surveillance. Thus, it is intractable for  UAVs to deal
with emergencies such as collision avoidance with other UAVs or obstacles
in time \cite{key-5}. Consequently, an effective surveillance system
for UAVs is essential to ensure flight safety. The primary issue for
surveillance is acquiring the positioning information. For instance,
radar is a candidate for detecting the UAV position \cite{key-6}.
However, there exist a couple of drawbacks by employing radar. In
detail, UAVs are too small to be detected by the radar, especially
in the bad weather \cite{key-7}. Besides, it is prohibitive or even
impossible to build sufficient radar stations due to the economic
and geographical factors. Compared with the radar, automatic dependent
surveillance-broadcast (ADS-B) is a competitive technique for the
future air traffic control (ATC) \cite{key-8}, and it is widely applied
in the field of civil aviation. ADS-B consists of two systems, ADS-B
IN and ADS-B OUT, which are responsible for receiving and broadcasting,
respectively. ADS-B helps avoid various flight occasions of UAVs.
For instance, in \cite{key-9}, a case of collision avoidance is provided
between  UAVs and helicopters. In addition, the trajectory prediction
for UAVs is an effective mechanism for surveillance. Based on the
data of UAVs, the flight trend can be figured out via predicting the
future trajectory, which provides useful information to avoid collisions
and supervise the UAV intrusion in the controlled airspace. 

The machine learning technique performs well when it meets the data
prediction, and it is widely used in the applications related with
the UAV \cite{key-10}. The recurrent neural network (RNN) and long
short-term memory (LSTM) specialize in capturing the characteristics
of data in the time dimension. In this case, when predicting time
series data, the two measures are outstanding \cite{key-11}. The
trajectory prediction for vehicles via machine learning becomes popular
in the research area. The UAV trajectory prediction in the airspace
is different with ground vehicles and vessels in the ocean. The vehicle
trajectory prediction benefits from the orderliness of the road traffic
network, which adds strong correlations to the neighbor trajectories.
Thus, the trajectory of vehicles is relatively easy to be predicted,
e.g., \cite{key-12} employs the social generative adversarial network
to predict the trajectory of automobiles. Due to the free path constraints,
the movement of vessels is random, and the prediction is intractable,
e.g., in \cite{key-13}, a novel sequence to sequence (seq2seq) model
is leveraged to predict the trajectory of vessels, which utilizes
the technique of gated recurrent unit (GRU) and LSTM. Different with
the two-dimensional movement of vehicles and vessels, UAVs  move freely
in the airspace, i.e., flying in a three-dimensional space with both
vertical and horizontal directions. Hence, it adds greater uncertainty
to the flight process of  UAVs and greater difficulty for the prediction
of the UAV trajectory. In addition, as for the civil aviation aircraft,
it has a long flight distance and a high flight altitude (about $10$
kilometers), so the changes of the trajectory directions are not arbitrary
and frequent. Besides, due to the heavy body, wind can hardly affect
the line of civil aviation aircrafts. However, since UAVs are small
and employed for special tasks, compared with the civil aviation aircraft,
the trajectory of UAVs changes significantly due to the task properties
or the influence of wind. The frequent and complex flight trajectories
bring significant challenges on predicting the trajectory. Based on
the neural relational inference, \cite{key-14} proposes a framework
for the two-dimensional trajectory prediction of  UAVs swarm. \cite{key-15}
employs 4 continuous trajectory points to predict future trajectory
data after the present time step via deep learning. The authors in
\cite{key-16} point out that, the prediction of the UAV trajectory
should consider extra information such as speed information, which
is an advantage of applying ADS-B on UAVs. 

Therefore, in this paper, the surveillance system of  UAVs is proposed
to deal with the conflicts detection in the low-altitude airspace
and emergency obstacle avoidance. ADS-B is adopted as the data source
of positioning information during the flight of UAV, which can figure
out the issue of insufficient trajectory data. Then, based on LSTM,
we propose the recurrent LSTM (RLSTM) to train and predict the three-dimensional
trajectory data for UAVs. 

The rest of this paper is organized as follows. The system model and
problem formulation are introduced in Section \ref{sec:System-Model}.
Then, the RLSTM based algorithm is presented in Section \ref{sec:Problem-Formulation}.
The simulation results are presented in Section \ref{sec:Simulation-Results}
and finally the conclusions are drawn in Section \ref{sec:Conclusions}.

\begin{figure}[t]
\centering

\includegraphics[width=3.5in]{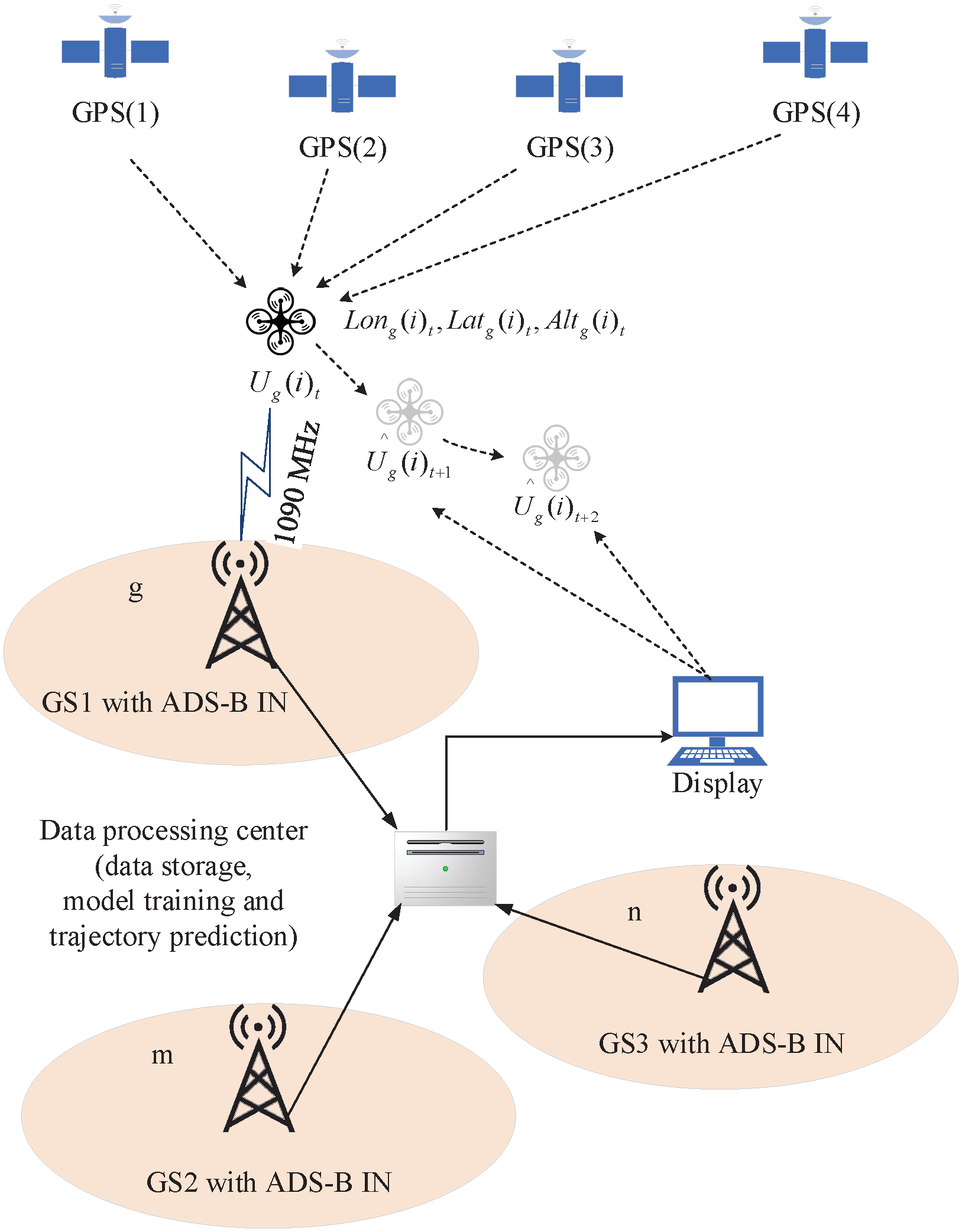}

\vspace{-2mm}

\textcolor{black}{\caption{\label{fig:Scenario}UAV surveillance system based on ADS-B information.}
}

\vspace{-5mm}
\end{figure}

\section{System Model and Problem Formulation\label{sec:System-Model}}

The ADS-B-based UAVs surveillance system is shown in Fig. \ref{fig:Scenario}.
The airspace is divided into several sub-airspaces, and each sub-airspace
has a corresponding ADS-B IN ground station ($GS$). The ground stations
transmit the received information to the data processing center. The
UAV can obtain its location from positioning information. The latitude
and longitude information ($Lat$, $Lon$) can been acquired from
the global positioning system (GPS) receiver via calculating the data
transmitted by four GPS satellites. The altitude information $Alt$
is obtained via the onboard pressure altimeter. When the airborne
ADS-B OUT system acquires the three dimensional (3D) positioning information,
it outputs the data according to a certain time interval. The broadcast
trajectory data are transmitted to the data processing center for
further monitoring and tracking. 

It is assumed that in the sub-airspace $g$, a set of UAVs $U_{g}$
($U_{g}(1)$, $U_{g}(2)$, $\ldots$, $U_{g}(n)$), which contains
$n$ UAVs and each UAV in $U_{g}$ broadcasts the positioning information
according to the pre-set broadcast interval. When UAV $U_{g}(i)$
employs ADS-B OUT in the $t$th broadcast time step, it requires the
positioning data ($Lat_{g}(i)_{t}$, $Lon_{g}(i)_{t}$, $Alt_{g}(i)_{t}$)
via the airborne GPS module and pressure altimeter. It is noted that
due to the lightweight and miniaturization of the current GNSS receiver
and pressure sensor (such as MPL3115A2 \cite{key-17}), the accurate
positioning for UAVs becomes available. When the positioning data
are obtained, the airborne ADS-B OUT system broadcasts the positioning
information to the airspace, and the information is received at the
corresponding ground station $GS1$. Since the information transmission
has time order, the trajectory of UAV $U_{g}(i)$ consists of a series
of flight trajectory data. Finally, the positioning information is
transmitted to the data processing center for storage, training and
prediction.

Trajectory prediction is an effective way to achieve surveillance
for UAVs, which can evaluate whether the future flight trend is safe,
and help avoid collisions with other UAVs or obstacles. The data processing
center employs the historical data of UAV $U_{g}(i)$ to predict the
future two trajectory points ($Lat_{g}(i)_{t+1}$, $Lon_{g}(i)_{t+1}$,
$Alt_{g}(i)_{t+1}$) and ($Lat_{g}(i)_{t+2}$, $Lon_{g}(i)_{t+2}$,
$Alt_{g}(i)_{t+2}$) at the broadcast time step $t$. The predicted
results provide a reference for the regulator to track and monitor
 UAVs, which guarantees the collision avoidance for all aircrafts
in airspace $g$.

During the process of UAV trajectory prediction, the prediction errors
are key indicators for evaluating the performance of different methods.
The flight of UAVs involves changes in longitude, latitude, and altitude.
Therefore, the prediction errors are divided into $Jx_{t}$, $Jy_{t}$,
$Jz_{t}$ and $J3d_{t}$, which refer to the prediction error of $Lat$,
$Lon$, $Alt$ and $3d$ at time step $t$. Besides, it is considered
that when the error of longitude and latitude is 1 degree, the error
of distance is about $114.1$ and $89.9$ kilometers, respectively.
Thus, the prediction errors are \vspace{-2mm}

\begin{equation}
Jx_{t}=\hat{Lat_{g}(i)_{t}}-Lat_{g}(i)_{t},\label{eq:(1)}
\end{equation}

\vspace{-2mm}

\begin{equation}
Jy_{t}=\hat{Lon_{g}(i)_{t}}-Lon_{g}(i)_{t},\label{eq:(2)}
\end{equation}

\vspace{-2mm}

\begin{equation}
Jz_{t}=\hat{Alt_{g}(i)_{t}}-Alt_{g}(i)_{t},\label{eq:(3)}
\end{equation}

\vspace{-2mm}

\noindent and

\vspace{-2mm}

\begin{equation}
J3d_{t}=\sqrt{\frac{Jx_{t}}{114,100}^{2}+\frac{Jy_{t}}{89,900}^{2}+Jz_{t}^{2}}.\label{eq:(4)}
\end{equation}

\noindent In formulas (\ref{eq:(1)})-(\ref{eq:(3)}), $\hat{Lat_{g}(i)_{t}}$,
$\hat{Lon_{g}(i)_{t}}$ and $\hat{Alt_{g}(i)_{t}}$ indicate the prediction
of $Lat_{g}(i)_{t}$, $Lon_{g}(i)_{t}$ and $Alt_{g}(i)_{t}$ at time
step $t$, respectively.

Since the predict process acquires two future trajectory points, according
to formulas (\ref{eq:(1)})-(\ref{eq:(4)}), the average prediction
errors $\bar{Jx_{t}}$, $\bar{Jy_{t}}$, $\bar{Jz_{t}}$ and $\bar{J3d_{t}}$
at time step $t$ are calculated as follows: \vspace{-2mm}

\begin{equation}
\bar{Jx_{t}}=\frac{Jx_{t+1}+Jx_{t+2}}{2},\label{eq:5}
\end{equation}

\vspace{-2mm}

\begin{equation}
\bar{Jy_{t}}=\frac{Jy_{t+1}+Jy_{t+2},}{2}\label{eq:6}
\end{equation}

\vspace{-2mm}

\begin{equation}
\bar{Jz_{t}}=\frac{Jz_{t+1}+Jz_{t+2}}{2},\label{eq:7}
\end{equation}

\vspace{-2mm}

\noindent and

\vspace{-2mm}

\begin{equation}
\bar{J3d_{t}}=\frac{J3d_{t+1}+J3d_{t+2}}{2}.\label{eq:8}
\end{equation}

The trajectory prediction for only once cannot evaluate the performance
of different algorithms. Consequently, continuous predictions for
UAV $U(i)$ for $n$ time steps are more convincing. Therefore, in
formulas (\ref{eq:9})-(\ref{eq:11}) we leverage the mean squared
error (MSE) $\bar{Jx}$, $\bar{Jy}$ and $\bar{Jz}$ as the average
prediction error: 

\vspace{-2mm}
\begin{equation}
\bar{Jx}=\frac{1}{n}\sum_{t=w}^{w+n}Jx_{t},\label{eq:9}
\end{equation}

\vspace{-2mm}
\begin{equation}
\bar{Jy}=\frac{1}{n}\sum_{t=w}^{w+n}Jy_{t},\label{eq:10}
\end{equation}

\vspace{-2mm}

\noindent and

\vspace{-2mm}
\begin{equation}
\bar{Jz}=\frac{1}{n}\sum_{t=w}^{w+n}Jz_{t}.\label{eq:11}
\end{equation}

Since the predicted trajectory points may perform different prediction
accuracy in different dimensions, i.e., good accuracy prediction and
poor altitude prediction. Hence, an accurate mechanism considering
various dimensions is required. Based on formulas (\ref{eq:9})-(\ref{eq:11}),
we choose the multi-dimensional MSE $\alpha$ as the loss function,
i.e., 

\vspace{-2mm}

\begin{equation}
\alpha=\frac{1}{3n}\sum_{t=w}^{w+n}(Jx_{t}^{2}+Jy_{t}^{2}+Jz_{t}^{2}).\label{eq:12}
\end{equation}

In order to achieve precise prediction accuracy, $\alpha$ in formula
(\ref{eq:12}) should be minimized. In Section \ref{sec:Problem-Formulation},
we propose the RLSTM and leverage the gradient back propagation to
deal with the minimization problem and predict future trajectory data.

\section{Algorithm Design\label{sec:Problem-Formulation}}

When the ground stations receive the ADS-B data from  UAVs, the data
are sequently transmitted to the data processing center. Therefore,
algorithms specialized in predicting time series data are considered
as effective candidates, such as the LSTM. LSTM is widely applied
to the natural language processing, since the predicted output of
LSTM is related to the input of current moment, and the state of the
hidden layer at the previous moment. ADS-B data have similar characteristics
with natural languages. For example, these data are generated in time
order, and there exist strong correlations between adjacent data.
LSTM has a unique structure to optimize the memory content. When new
data come, the \textquotedbl{}memory gate\textquotedbl{} and \textquotedbl{}forgetting
gate\textquotedbl{} determine which information should be recorded
into the cell state, and which information should be forgotten. The
cell state updates much slower than the hidden state. As above, LSTM
is a competitive method for the UAV trajectory prediction.

\begin{algorithm}[t]
\caption{\label{Algorithm1-1} RLSTM for UAV trajectory prediction.}

\begin{algorithmic}[1]

\REQUIRE Training epoch $e$, training iteration $k$, training set
$T$, predicted trajectory $U$, learning rate $lr$ and the trajectory
data received time step $t$.

\ENSURE Predicted trajectory data: $p_{t+1}$ and $p_{t+2}$.

\STATE\textit{Initialization: }$e=2$, $k=300$, $lr=0.01$ and $t=0$.

\FOR {$i=1$ to $e$} 

\FOR {$j=1$ to $k$} 

\STATE Leverage LSTM to train the model with training set $T$.

\STATE Update the weights of model.

\ENDFOR

\ENDFOR

\REPEAT

\STATE Receive new trajectory point: $t=t+1$.

\WHILE {$t>16$} 

\FOR {$f=1$ to $k$}

\STATE Leverage LSTM to train the model with data from $1$st to
$t$th.

\ENDFOR

\STATE Leverage continuous data from $(t-15)$th to $t$th to predict
the next two trajectory points.

\RETURN $p_{t+1}$, $p_{t+2}$.

\ENDWHILE

\UNTIL $t$ is the last time step of the predicted UAV trajectory.

\end{algorithmic}
\end{algorithm}

Based on the advantages of LSTM, we further propose the RLSTM algorithm.
The entire neural network consists of an LSTM network and a fully
connected network. If there exist no data similar to the new data
in the training set, it brings great errors for the prediction results.
In order to obtain the UAV\textquoteright s trajectory prediction
refer to the historical data and dynamically predict the new trajectory
with high precision, we propose the RLSTM, which adopts a recurrent
training-prediction structure. Based on learning the data features
from the training set, the model is retrained with all the historical
data, which belong to this predicted trajectory at the current moment
during each prediction process. Then, the new model is leveraged to
make prediction for future two time steps.

Due to the limitation of energy capacity, the flight distance of  UAVs
is limited, which leads to the latitude and longitude data varying
in a small range. Thus, it results in slow gradient descent when training
the model with LSTM, and causes negative effects. To deal with this
issue, we can preprocess the ADS-B data before training the model.
The RLSTM leverages Z-score normalization for processing the trajectory
data. In the following formulas, $\mu$ and $\sigma$ are the mean
and variance of the data, respectively. The processed data have a
mean of $0$ and a standard deviation of $1$. The positioning data
normalized by Z-score at time step $t$ are 

\vspace{-2mm}

\begin{equation}
Lat_{g}(i)_{t}^{*}=\frac{Lat_{g}(i)_{t}-\mu_{Lat_{g}(i)}}{\sigma_{Lat_{g}(i)}},\label{eq:13}
\end{equation}

\vspace{-2mm}

\begin{equation}
Lon_{g}(i)_{t}^{*}=\frac{Lon_{g}(i)_{t}-\mu_{Lon_{g}(i)}}{\sigma_{Lon_{g}(i)}},\label{eq:14}
\end{equation}

\vspace{-2mm}

\noindent and

\vspace{-2mm}

\begin{equation}
Alt_{g}(i)_{t}^{*}=\frac{Alt_{g}(i)_{t}-\mu_{Alt_{g}(i)}}{\sigma_{Alt_{g}(i)}}.\label{eq:15}
\end{equation}

\noindent In formulas (\ref{eq:13})-(\ref{eq:15}), $Lat_{g}(i)_{t}^{*}$,
$Lon_{g}(i)_{t}^{*}$ and $Alt_{g}(i)_{t}^{*}$ refer to the normalized
positioning data $Lat_{g}(i)_{t}$, $Lon_{g}(i)_{t}$ and $Alt_{g}(i)_{t}$,
respectively.

Firstly, the previous trajectories of  UAV which are preprocessed
and leveraged serve as the training set. When the training process
is finished, fifteen consecutive trajectory data are selected randomly
as the training data for one round of recurrent training-prediction
process, to calculate the loss function and update the entire model
according to the learning rate. When new trajectory data are transmitted
to the data processing center, recurrent training-prediction process
is repeated. The RLSTM is described in detail as follows. 

When the ADS-B data of  UAVs are preprocessed, the RLSTM runs according
to Algorithm \ref{Algorithm1-1}. To begin with, the model training
is performed on UAVs training set. The training set consists of different
UAV trajectories in the current airspace, and each one has several
trajectory points. We set the epoch of the entire training set $e$
as $2$ and count the number of trajectory data as $t$. The algorithm
begins when $t>16$, since $17$ consecutive trajectory data are the
minimum requirements for starting recurrent training-prediction progress.
In each epoch, every trajectory is trained for $300$ times, and in
each iteration, the first data $x$ is randomly selected as the starting
trajectory point. Then, employ the $x$th to $(x+15)$th as the training
data for updating the model. We leverage the trained model to predict
the trajectory points after $2$ time steps. The $(x+16)$th and $(x+17)$th
are used to calculate the loss function and update the weights of
the model according to the learning rate. Then, we conduct recurrent
training-prediction process. For example, when the 17th data comes,
the model training is activated. The RLSTM uses the $1$st to $15$th
trajectory points as training data and trains for $300$ times. When
the training process is finished, formula (\ref{eq:12}) is used to
calculate the loss function. We leverage the Adam optimizer to optimize
the gradient descent. The algorithm updates the cell state, hidden
state and network weight, and sends the $2$nd to $17$th trajectory
points into the model to predict the $18$th and $19$th trajectory
points. When receiving the two trajectory points, formula (\ref{eq:8})
calculates the average prediction error. This process repeats until
there is no new data received. When the training process is completed,
the model is saved for further prediction.

\section{Simulation Results\label{sec:Simulation-Results}}

\begin{figure}[t]
\centering

\includegraphics[width=3.5in]{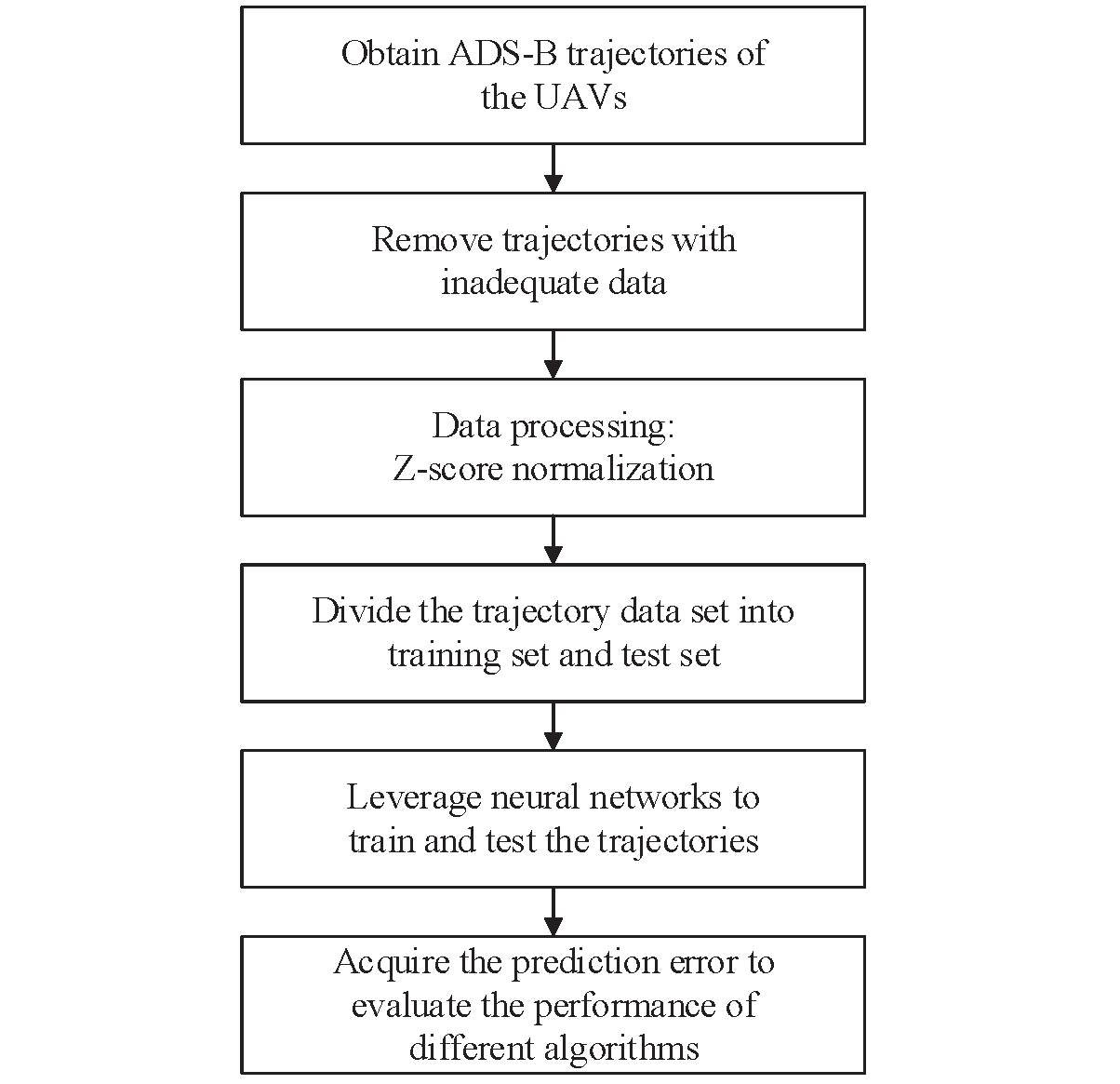}

\vspace{-3mm}

\caption{Process of trajectory prediction.\label{fig:2}}

\vspace{-5mm}
\end{figure}

The simulation process is shown in Fig. \ref{fig:2}. The multiple
existing UAVs trajectory data set is the premise for subsequent simulations.
The trajectories with inadequate data are removed from the data set,
since they may lead to large prediction error. When the trajectory
data are normalized, the data set is divided into training set and
test set, and we employ different neural networks to train and predict
the dataset and finally obtain the prediction error.

In order to evaluate the performance of the RLSTM in trajectory prediction
for the proposed UAV surveillance system equipped with ADS-B, we select
multilayer perceptron (MLP), RNN, LSTM and bi-directional LSTM (Bi-LSTM)
as the comparison algorithms. MLP consists of $3$ layers with fully
connected network, and the quantity of weights in each fully connected
layer is $100$. RNN, LSTM and Bi-LSTM consist of a hidden layer and
a fully connected layer. The hidden-size is set as $16$. The parameters
are set as: the training epoch and iteration we set are $2$ and $300$,
respectively. The trajectories of UAVs in the training set are $67$,
and the trajectories in the test set are $7$. Trajectory points in
the test set are larger than $25$, and larger than $20$ in the training
set. In order to make the simulation results more convincing, the
prediction is repeated for $10$ times. In each prediction process,
the prediction error is the average of the two predicted trajectory
points and the real data received.

\begin{figure}[t]
\centering

\subfloat[Predicted trajectory of T1\label{fig:sub-1}]{\includegraphics[width=3.5in]{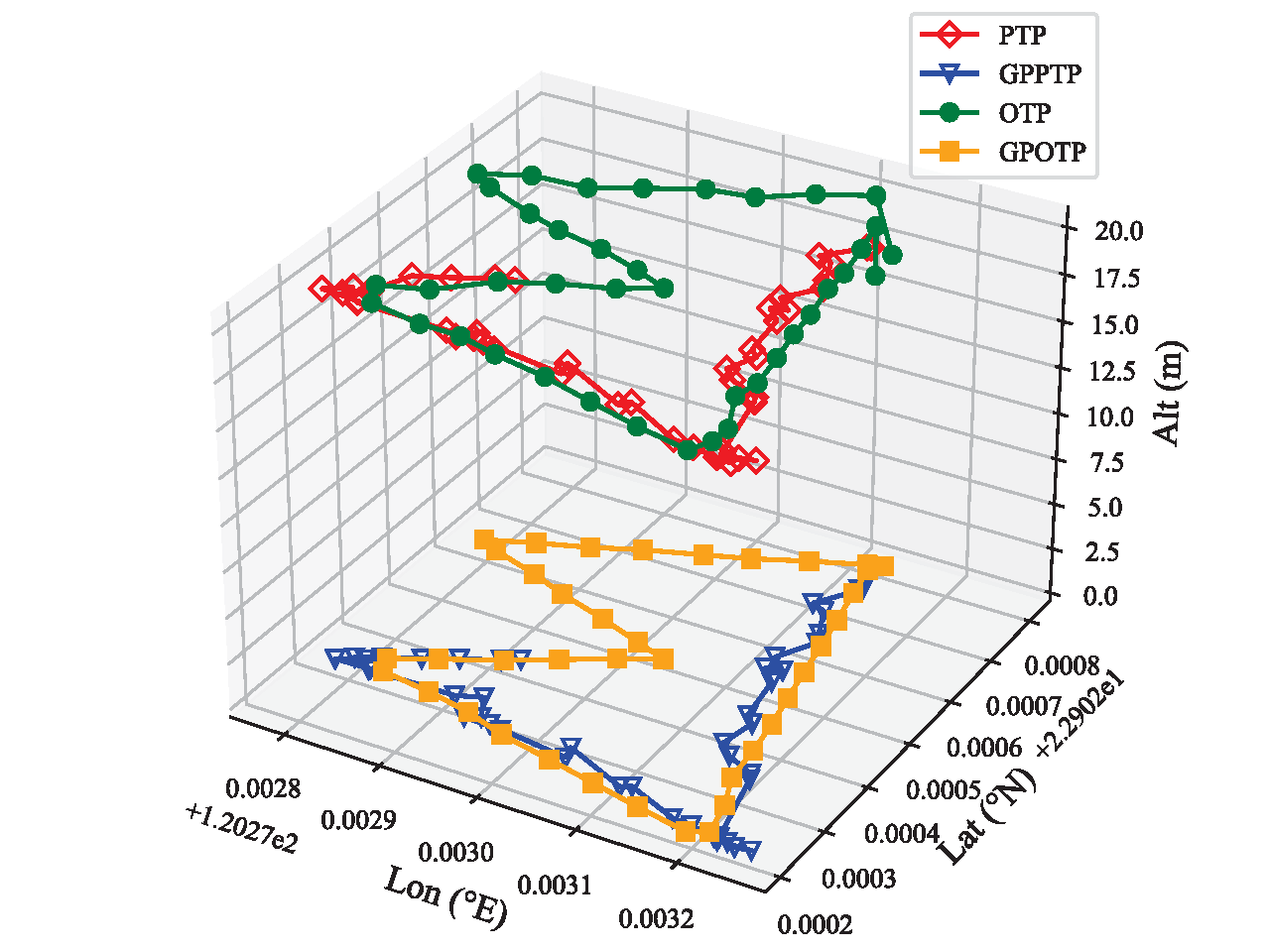}

\vspace{-3mm}

}

\vspace{-3mm}

\subfloat[Average prediction error of T1 in 3D.\label{fig:sub-2} ]{\includegraphics[width=3.5in]{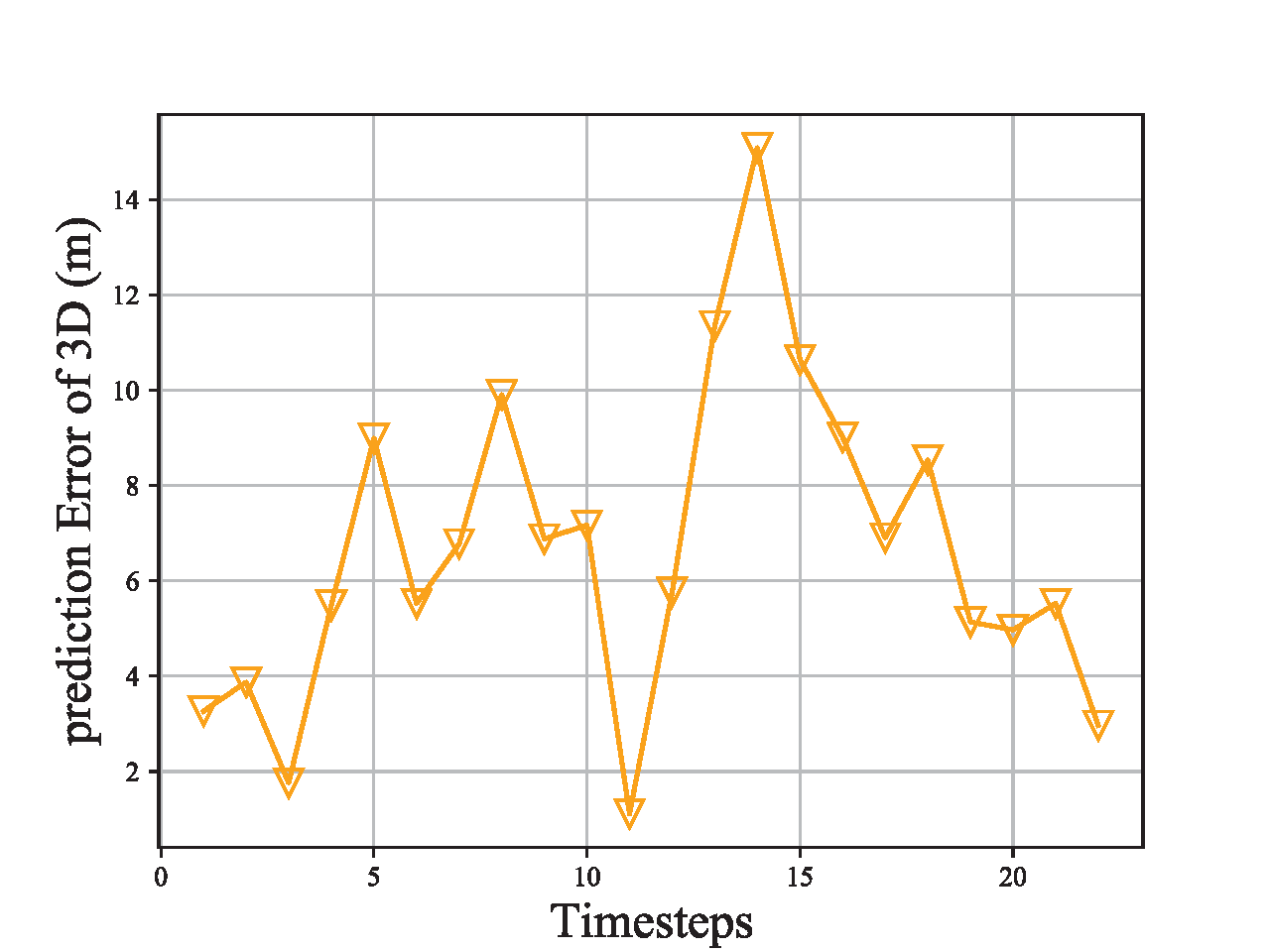} 

\vspace{-3mm}

}

\caption{Predicted trajectory and average prediction error in 3D via employing
the RLSTM.\label{fig:3.1}}

\vspace{-5mm}
\end{figure}

In the simulation, the flight positioning data are chosen from Kaggle
\cite{key-18}, which is broadcast by the ADS-B OUT system in the
surveillance system. The broadcast interval of each trajectory point
is about $2s$-$3s$. Frequent update of ADS-B information leads to
the congestion and interfere. Hence, it is unacceptable for the aviation
aircrafts. Since the flying speed of  UAVs is much slower than civil
aviation aircraft, it is not necessary for the UAV to frequently update
the ADS-B information. In short, the selection of data set is suitable
for our proposed surveillance system.

\begin{figure}[t]
\centering

\includegraphics[width=3.5in]{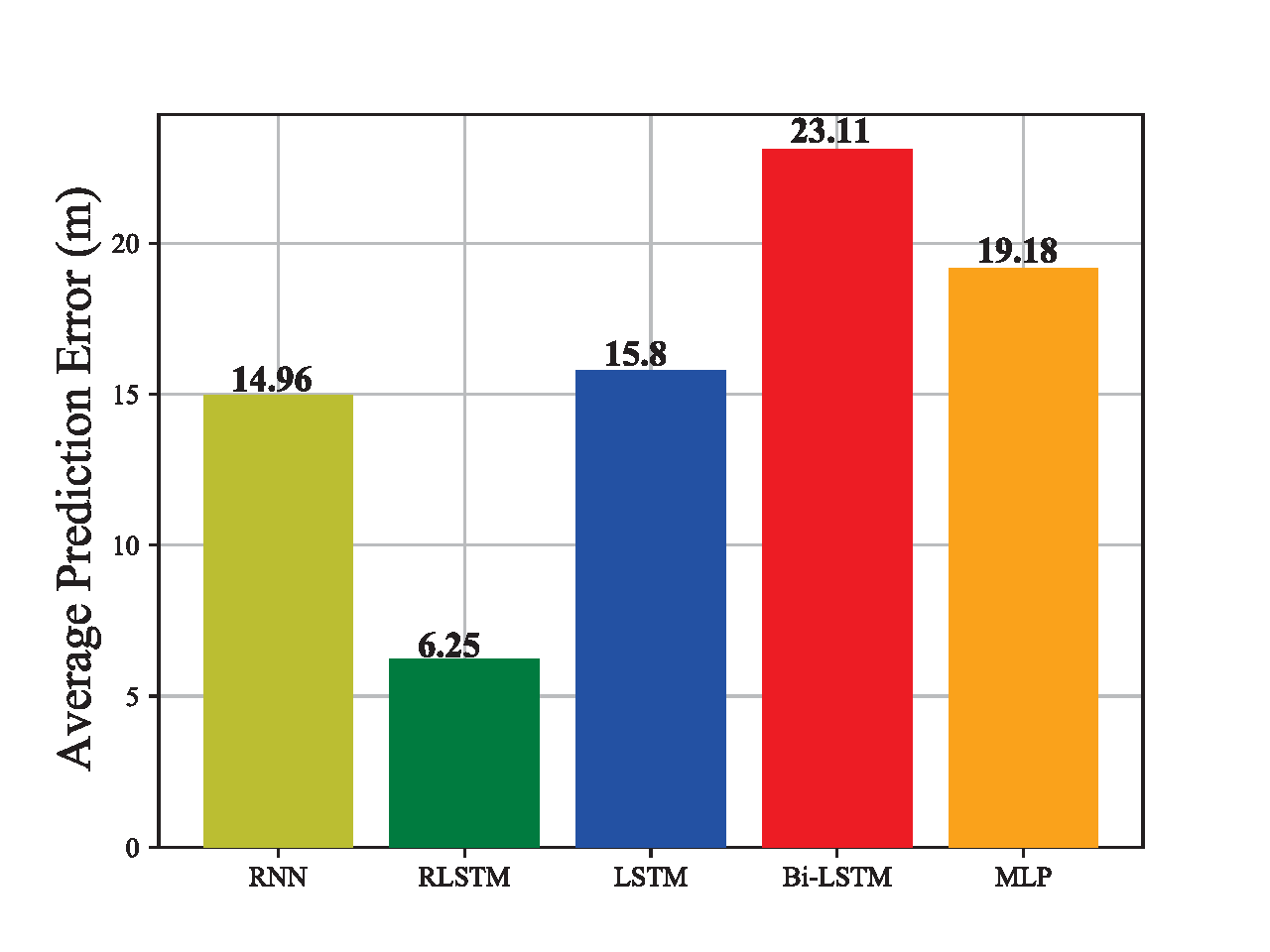}

\vspace{-3mm}

\caption{Average prediction error of different algorithms.\label{fig:4.1} }

\vspace{-5mm}
\end{figure}

\begin{figure}[t]
\centering

\includegraphics[width=3.5in]{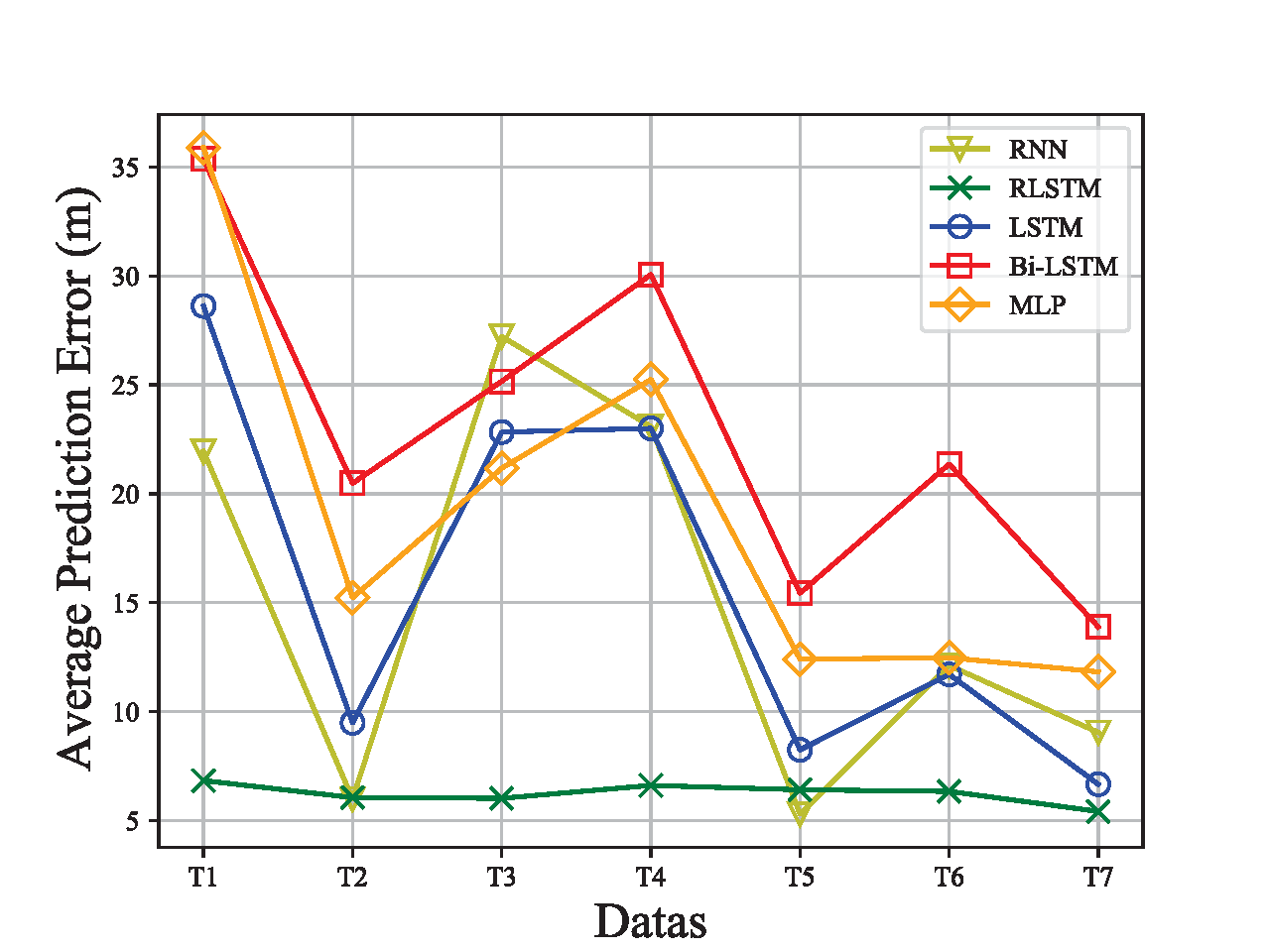}

\vspace{-3mm}

\caption{Average prediction error of each test data via employing different
algorithms.\label{fig:5error} }

\vspace{-5mm}
\end{figure}

Fig. \ref{fig:3.1} shows the complete prediction process of the trajectory
$T1$ in the test set via leveraging the RLSTM, and the trajectory
direction is counterclockwise. As validated in Fig. \ref{fig:sub-1},
the first $16$ points in the original trajectory point (OTP) are
not employed for prediction. When the $17$th point is received, the
predicted trajectory points (PTP) of $18$th and $19$th time steps
are predicted. The subsequent processes are made according to Algorithm
\ref{Algorithm1-1}. In order to observe the latitude and longitude
variation, the ground projection of OTP (GPOTP) and ground projection
of PTP (GPPTP) are validated in Fig. \ref{fig:sub-1}. In Fig. \ref{fig:sub-2},
the average error of $T1$ for each prediction time step is presented,
according to formula (\ref{eq:8}). The reason for the rise of prediction
error at $3$rd to $8$th time steps is due to that the direction
of the UAV trajectory changes at this time period. The time steps
$11$ to $14$ are the second surge of prediction error, and it is
attributed to the change of trajectory. The $17$th to $21$st time
steps vary since the predicted trajectory is continuously revised,
and the prediction error drops to an acceptable range finally.

Fig. \ref{fig:4.1} is the average error according to $10$ repeated
predictions on the test set data via different algorithms. A low average
prediction error refers to satisfied performance. It is validated
from Fig. \ref{fig:4.1} that the RLSTM performs better in predicting
trajectory points with a prediction error of $6.25$ meters. The prediction
error on RNN, which is the suboptimal prediction accuracy is reduced
to $14.96$ meters, and it is also $8.71$ meters smaller than the
results of RLSTM.

Fig. \ref{fig:5error} reveals the average prediction error of each
trajectory in test set for repeating prediction $10$ times. Obviously,
the average prediction error of the RLSTM is lower than other compared
algorithms. It is worth noting that for the trajectory at $T1$, $T3$,
and $T4$, large average prediction errors appear in all considered
algorithms. RLSTM is the only algorithm, which is able to effectively
control the error within $10$ meters compared with the other algorithms.

It is confirmed from the simulation results that, compared with other
algorithms, the RLSTM can achieve higher accuracy of trajectory prediction
in the proposed UAVs surveillance system equipped with ADS-B.

\section{Conclusions\label{sec:Conclusions}}

In this paper, we propose a system for UAV surveillance based on employing
ADS-B as the positioning data source. In order to reach high performance
of the proposed system, we put forward the RLSTM to repeat the recurrent
training-prediction process on the trajectory, and acquire the predicted
two trajectory points for the next time steps. The simulation results
reveal that, compared with other neutral networks, the RLSTM performs
better. The prediction error of the RLSTM is about 6 meters in average,
which is within the range of availability for UAVs in the airspace.

\end{document}